\documentclass[authorversion,nonacm, screen]{acmart}
\usepackage{enumerate}

\AtBeginDocument{%
  }

\begin{document}

\title{Evolving AI Risk Management: A Maturity Model based on the NIST AI Risk Management Framework}


\author{Ravit Dotan}
\email{eravit@gmail.com}
\affiliation{%
  \institution{TechBetter}
    \country{USA}
}

\author{Borhane Blili-Hamelin}
\email{borhane@avidml.org}
\affiliation{%
  \institution{AI Risk and Vulnerability Alliance}
  \city{Brooklyn}
  \state{New York}
  \country{USA}
}

\author{Ravi Madhavan}
\email{ravi.madhavan@pitt.edu}
\affiliation{%
  \institution{The University of Pittsburgh}
    \country{USA}
} 
\author{Jeanna Matthews}
\email{jnm@clarkson.edu}
\affiliation{%
  \institution{Clarkson University}
    \country{USA}
}  

\author{Joshua Scarpino}
\email{scarpinojosh@gmail.com}
\affiliation{%
  \institution{TrustEngine}
  \institution{Assessed.Intelligence}
    \country{USA}
}

\renewcommand{\shortauthors}{Dotan et al.}

\begin{abstract}
Researchers, government bodies, and organizations have been repeatedly calling for a shift in the responsible AI community from general principles to tangible and operationalizable practices in mitigating the potential sociotechnical harms of AI. Frameworks like the NIST AI RMF embody an emerging consensus on recommended practices in operationalizing sociotechnical harm mitigation. However, private sector organizations currently lag far behind this emerging consensus. Implementation is sporadic and selective at best. At worst, it is ineffective and can risk serving as a misleading veneer of trustworthy processes, providing an appearance of legitimacy to substantively harmful practices. In this paper, we provide a foundation for a framework for evaluating where organizations sit relative to the emerging consensus on sociotechnical harm mitigation best practices: a flexible maturity model based on the NIST AI RMF. 
\end{abstract}




\maketitle

\section{Introduction}

In recent years, increasingly more professionals in the AI ethics space have been calling for “operationalizing AI ethics” or “translating principles into practice” –  meaning moving away from articulating general priorities and principles, which has been prominent in the last decade, into establishing processes that rigorously anticipate, evaluate, mitigate, and provide redress for AI harm \cite{drew_design_2018, morley_what_2020, morley_operationalising_2023, ayling_putting_2022, zhu_ai_2021, munn_uselessness_2023}. On that backdrop, practitioners, researchers, and government bodies have developed recommendations and practices to bridge the gap \cite{holland_dataset_2018, gebru_datasheets_2021, mitchell_model_2019, nist_airmf_2023}.

Despite the existence of a host of AI ethics frameworks and tools, organizations have been lagging behind the recommended best practices \cite{ibm_ibm_2022, mckinsey_state_nodate, dotan_evaluating_2024}. For example, McKinsey \cite{mckinsey_state_nodate} shows that in 2022 only 17\% of companies reported they worked to mitigate fairness and bias issues, merely a small increase from the 13\% who did so in 2019, despite the many tools and frameworks that were developed in between. Dotan et al. \cite{dotan_evaluating_2024} show that, while 76\% of large companies reported making AI ethics commitments, employing AI ethics personnel, or engaging in thought leadership in 2022, only a small fraction reported engaging in AI risk management best practices such as data and model documentation (9\%) and maintaining an incident log (2\%).

This paper presents a foundation for a maturity model that is intended to help companies decrease this gap. Maturity models guide companies by laying a sequence of stages for progress and are widely used in many areas, from cybersecurity to software development best practices \cite{poppelbus_what_2011}. A maturity model grounded in AI ethics could help organizations evaluate their existing AI risk management practices and plan how to do better \cite{vakkuri_time_2021}.

We chose to base the maturity model we describe in this paper on the NIST AI Risk Management Framework (AI RMF)\cite{nist_airmf_2023} for practical reasons and because of an alignment with the values embodied in the AI RMF. We argue for practical advantages of basing AI ethics maturity models in widely accepted frameworks, and we explain why we favored the RMF over the EU AI Act. Moreover, we highlight the fact that designing a maturity model requires making substantive value decisions because of the inherent evaluative aspect \cite{bommasani_evaluation_2022}. We chose the NIST AI RMF for its focus on sociotechnical harm and its elevation of marginalized voices through numerous design choices. 

The structure of the paper is as follows. In Section 2, we provide a brief background of maturity models and their purpose, highlight the current maturity model landscape within AI, and provide a more detailed overview of the NIST AI RMF and our reasons for aligning our maturity model with it. In Section 3, we describe the questionnaire leveraged within our maturity model, and in Section 4, we explain the scoring approach we use, as well as a detailed description of the components of the scoring evaluation process. Section 5 is focused on explaining maturity model score aggregation. Section 6 covers research limitations and the next steps. The full questionnaire and an example of how it works in practice are in the appendices.

\section{Background}

\subsection{About maturity models}

A staple of current technology management toolkits, maturity models are ''conceptual multistage models that describe typical patterns in the development of organizational capabilities” \cite{poppelbus_what_2011}. They have been characterized as a Crawl/Walk/Run-style set of factors depicting the progression of capabilities while also serving as a tool to benchmark current capabilities and help set goals and priorities for improvement \cite{c2m2_us_department_of_energy_cybersecurity_2022}. The practical utility of maturity models stems from their simplicity, conceptual power, and evolutionary orientation, which result in effective managerial guidance on where to invest attention, effort, and other resources in order to build capability in successive stages (see Poeppelbus et al. \cite{poppelbus_what_2011} and \cite{wendler_maturity_2012} for overviews of the large body of literature on maturity models, and Poeppelbuss \& Röglinger \cite{poppelbus_what_2011} for a discussion of design principles for maturity models).

The use of maturity models in technology management dates back to the 1980s, with predecessors dating back to the 1960s \cite{piaget_cognitive_1964, kuznets_economic_1965, crosby_quality_1979, proenca_risk_2017}. As the popularity of maturity models has increased, their application has spread to many capability arenas. Well-known examples include the Software Capability Maturity Model \cite{CMMI_cmmi_nodate} and the risk management maturity model \cite{proenca_risk_2017}. NIST-related maturity models include the NIST cybersecurity maturity model (National Cybersecurity Center of Excellence \cite{nist_cybersecurity_2023}), and the simple Privacy maturity model in NIST's privacy framework in the form of “Ready, Set, Go” labels \cite{nist_privacy_2020}. Moreover, NIST worked with the US Department of Defense (DOD) to help create Cybersecurity Maturity Model Certification (CMMC)\cite{ccmc_cmmc_2021} and with the Department of Energy (DOE) to produce mappings between the NIST 800-53 and 800-171, the Cybersecurity Framework (CSF) and Cybersecurity Capability Maturity Model (C2M2) \cite{c2m2_us_department_of_energy_cybersecurity_2022} \cite{nist_cybersecurity_2023}.

Maturity models have great potential in AI ethics, too. They can help companies understand where they stand with regard to standards, analyze gaps, and plan for improvement \cite{vakkuri_time_2021}. This support can help overcome the low implementation of AI ethics, which, as discussed above, is prevalent and can result in a great deal of harm to individuals, society, and the companies themselves. 

\subsection{Maturity models in AI ethics}
A handful of maturity models already exist in AI ethics, most of which were created by private sector companies, especially big tech and consulting firms. Those offered by Salesforce \cite{baxter_k_ai_2021} and Microsoft \cite{microsoft_vorvoreanu_responsible_2023} may be the most well-known. Other maturity models dedicated to AI ethics include ODI \cite{open_data_institute_data_2022}, Ethical Intelligence \cite{ethical_intelligence_ethics_2022}, and Krijger et al. \cite{krijger_ai_2023}. Moreover, some maturity models for the development of AI capabilities, such as IBM \cite{ibm_ai_2021},  MITRE \cite{mitre_mitre_2023}, and PwC \cite{pwc_responsible_2021}, include maturity in AI ethics as an aspect of maturity in AI adoption. 

The existing AI ethics maturity models are limited. To start, they assume specific trajectories toward good AI governance, while in reality, there may be multiple legitimate trajectories, especially in different contexts. Rigid trajectory expectations may render a model unhelpful or even misleading when used in the wrong context. 

One trajectory expectation, for example, is that buy-in from senior management comes at the last stage (e.g. Microsoft \cite{microsoft_vorvoreanu_responsible_2023}). This assumption sounds reasonable in large corporations such as Microsoft. However, in small to medium-sized enterprises (SMEs), research shows that the motivation for responsible innovation typically comes from the founder, so C-suite buy-in comes first (\cite{Covello&Iatridis}, \cite{Bos-Brouwers}). Another expectation that AI governance starts with policy making that later develops into implementation (e.g., Salesforce \cite{baxter_k_ai_2021}, ODI \cite{open_data_institute_data_2022}). Again, this expectation may sound reasonable in large corporations such as Salesforce. However, in SMEs, which tend to be more informal, company-wide policies may only develop later in the process. A startup we got feedback from raised this concern explicitly, worrying that the fact that their AI ethics processes have not been formalized into written company-wide policies will overshadow the fact that they are implementing AI ethics processes regardless. Other models assume a trajectory in which teams start with localized activities that are later generalized into company-wide policies \cite{krijger_ai_2023}). While this trajectory may be more amenable to smaller companies that allow leeway to individual teams, it may be a poor fit for a large corporation. 

All of these different trajectories may be fruitful in the right context, and AI ethics maturity models should be inclusive of multiple trajectories fitting to a variety of contexts. The existing models lack this feature. As we will show, the maturity model presented in this paper is open to multiple maturity trajectories. 

Another limitation of the existing AI ethics maturity models is that they are based on bespoke conceptual frameworks that they develop, including their own dimensions of progress and maturity stages along the dimensions. Microsoft’s \cite{microsoft_vorvoreanu_responsible_2023} model, for example, has five maturity stages (Latent, Emerging, Developing, Realizing, and Leading) and five dimensions of progress (Organizational foundations, Team approach, Cross-discipline collaboration, and Responsible AI practices). For comparison, Ethical Intelligence’s \cite{ethical_intelligence_ethics_2022} model presents three maturity stages (Below, Average, and Exemplary) and five dimensions (Accountability, Social impact, Intentional design, Trust and transparency, and Fairness). 

The bespoke approach reinvents the wheel and fails to take advantage of the conclusions, expertise, terminology, or authority of widely accepted documents and initiatives. This creates practical challenges. For one thing, reinventing the wheel with new terminology or a custom framework increases the complexity of adoption, and it makes it difficult for companies to communicate their results to others and benchmark their results relative to other organizations. Moreover, bespoke approaches are difficult to relate to industry standards and regulatory proposals. Therefore, even when an organization meets the expectations of a bespoke maturity model, there will likely still be confusion about whether it meets the expectations of the industry or regulators. 

Further, the reliance on bespoke frameworks is also problematic due to the political undertone of AI ethics maturity models, especially in light of the fact that many of the existing models were created by technology companies themselves. By nature, maturity models guide organizations to achieve some ''good,” in this case, good governance of AI. Defining what counts as ''good,” especially in the case of AI governance, is not only thorny but also deeply political. In a narrow sense, defining what counts as good AI governance is political because various organizations, in particular big tech, are lobbying regulators to shape regulation about AI governance in ways that are favorable to them \cite{Oprysko}. As illustrated in the previous paragraph, what is favorable to big tech may be unfavorable to SMEs, not to mention other stakeholders such as the public. In a wider sense, AI governance is political because it involves decision-making about topics that are political, such as expectations around the protection of civil and human rights. Due to this political nature, it matters who gets to determine what counts as ''good” AI governance. Big tech and other private sector organizations may fail to represent the public good in their AI ethics maturity models due to conflicts of interest or a more narrow view of what risks to prioritize in risk management.

\subsection{NIST AI RMF}
The NIST AI RMF, the basis of our maturity model, is a voluntary framework describing best practices for AI risk management, including concrete activities for the development and deployment of AI in a socially responsible way. It is one of the most well-respected documents on AI governance and is growing in influence, especially in light of the October 2023 Executive Order on the Safe, Secure, and Trustworthy Development and Use of Artificial Intelligence that specifically calls out the NIST AI RMF \cite{the_white_house_executive_2023}. The many AI companies based in the United States may view the US-based policies as especially relevant.

The RMF's influence is one of the primary reasons we chose to use it as the basis of our maturity model, as we seek to avoid the challenges of bespoke frameworks discussed above.

In addition to the NIST AI RMF, we considered the creation of a maturity model with respect to the European Union’s AI Act, but chose NIST’s AI RMF for two primary reasons. First, many of the governance requirements in the EU AI Act are focused on high-risk systems, and companies that don’t fall into this bucket may argue that it is not relevant to them. We wanted a maturity model that was not specifically limited to high-risk systems and could be leveraged for all AI systems. Second, we wanted to encourage viewing maturity as a continually evolving lifecycle and not simply a checklist for compliance with regulation. The voluntary nature of the NIST AI RMF allows organizations to view maturity of AI risk management in this way. 

Another reason we chose the NIST AI RMF as the basis of our maturity model is its emphasis on a sociotechnical perspective on AI harm. Over the past 10 years, researchers and organizations have converged on the importance of relying on a plurality of sociotechnical methods to map, measure, disclose, and manage the interlocking technical and social causes of algorithmic harms.  \cite{selbst_fairness_2019, weidinger_sociotechnical_2023} Moreover, grappling seriously with the interlocking technical and social causes of AI harm requires centering the perspectives of stakeholders impacted by AI systems, rather than merely of stakeholders involved in building and deploying those systems \cite{lazar_ai_2023}. The AI RMF explicitly and repeatedly calls upon organizations to embrace this kind of sociotechnical perspective on AI risk management \cite{nist_airmf_2023}.

Last, the RMF naturally lends itself to a maturity model, and a maturity model based on it complements it. First, the RMF lends itself to a maturity model because the RMF touches directly on many high-level activities that are relevant to maturity, such as measuring risk and internal policy making. Their focus on activities throughout the lifecycle of an AI system makes it an especially good candidate for a maturity model. Second, our maturity model complements the RMF because the RMF is not intended to provide guidance on how organizations might evolve towards the best practices being recommended or on how to evaluate the extent to which organizations are aligned with those best practices. Therefore, organizations may struggle with how to prioritize plans for improvement. Similarly, external actors, such as investors and consumers, may struggle to use the NIST AI RMF to evaluate organizations and their level of maturity with respect to implementing the RMF. In fact, one could wonder why the NIST AI RMF does not itself contain a maturity model, especially given that NIST has developed explicit maturity model language for other related areas such as cybersecurity and privacy. The creation of a maturity model will assist in the transition from ad hoc implementation of responsible AI to more mature processes and programs \cite{vakkuri_time_2021}.  

\subsection{Design process}

We apply the maturity model approach to translate the AI risk considerations identified in the RMF to an evolutionary description. Our work unfolds in two steps: an inductive step to lay out the foundation of the maturity model, which we do in this paper, and a confirmatory step with case studies and empirical refinement, which we do in ongoing and future work, described in the last section, Section \ref{limitations-section}.

This approach differs from current models within AI ethics and addresses challenges encountered in the field of maturity model design. Maturity models have been critiqued for oversimplification and inadequate empirical grounding. Accordingly, an overarching current concern in the field is how to develop maturity models that are both theoretically robust and empirically grounded \cite{poppelbus_what_2011}. In our case, the conceptual strength of our framework comes from the conceptual strength of the RMF and the approach developed in this paper. The empirical strength will come from the empirical work based on this foundation. 

In developing the conceptual framework in this paper, the inputs we used were the NIST AI RMF and the broader literature on the development of maturity models, leading to the initial specification of the phases and content of the maturity model. The model was iteratively refined by the authors to reflect the appropriate stages of, and linkages between, the RMF elements over the development of AI capability. The refined model was then pilot-tested through consultations and dry runs with a convenience sample of AI stakeholders, including startup leaders, active investors and AI ethics experts, including NIST team members involved in developing the RMF. Throughout this process, the model was successively refined via the incorporation of feedback with regard to accuracy, clarity, relevance and ease of use. The model presented in this paper represents the current stage of its development based on that process.

\section{Flexible Questionnaire}

Our maturity model includes a flexible questionnaire and scoring guidelines. The questionnaire consists of a list of statements, and evaluators are asked to rank them using the scoring guidelines discussed below. The statements in the questionnaire center on concrete and verifiable actions, such as conducting certain processes and documenting the outcomes. For example: 
\begin{quote}
    ''We regularly evaluate and document bias and fairness issues caused by our AI systems”. 
\end{quote}

The questionnaire avoids general and abstract statements such as ''Our AI systems are fair”. Further, the statements use the plural first pronoun ''we” and the active present tense, e.g., ''we document.” This is an intentional choice made to emphasize the responsibilities of the companies and people who manage AI. 

The statements cover the content of the RMF's governance recommendations, which are divided into four pillars: MAP - Learning about AI risks and opportunities; MEASURE - Measuring risks and impacts; MANAGE - Implementing practices to mitigate risks and maximize benefits; and GOVERN - Systematizing and organizing activities across the organization. Each of the pillars includes a list of categories and subcategories. For example, one of the categories in the MEASURE pillar is ''MEASURE 2: AI systems are evaluated for trustworthy characteristics.” One of the subcategories in this category is ''MEASURE 2.11: Fairness and bias – as identified in the MAP function – are evaluated and results are documented.” \cite{nist_airmf_2023}. In isolation, each statement in the questionnaire covers one or more of the NIST AI RMF subcategories. For example, the statement above covers the subcategory MEASURE 2.11. Jointly, the statements in the questionnaire cover all RMF subcategories. 

The questionnaire is flexible in that evaluators are not required to evaluate all statements. The questionnaire allows the evaluator to adjust the evaluation to the organization’s specific context in three key ways: 1) level of granularity, 2) life-cycle stage of the AI system, and 3) multiplicity of AI systems within the organization. We elaborate on each of these in the subsections that follow.

\subsection{Flexibility 1: Granularity}
Evaluators who are interested in a fine-grained evaluation can rank each of the 60 statements in the questionnaire. However, the feedback we received from practitioners indicates that many are interested in a more coarse-grain evaluation. Therefore, the 60 statements are divided into 9 topics. Each topic is represented by a sentence that describes the statements in that topic. For example, one of the topics is
\begin{itemize}
    \item Topic 4 - ''\emph{Measuring risk:} We measure our potential negative impacts”.    
\end{itemize}

Under this topic, there are finer grain individual statements, including for example:
\begin{itemize}
    \item Statement 4e:  ''We regularly evaluate and document bias and fairness issues related to our AI systems”. 
    \item Statement 4i: ''We regularly evaluate and document security issues related to our AI systems”.
\end{itemize}
    
Those interested in a coarse-grained evaluation can score only the topic statement. However, the individual statements are still taken into account because, as discussed in more detail below, the scoring guidelines instruct evaluators to give higher scores the better the coverage of the individual statements. 

\subsection{Flexibility 2: Life-cycle stage}

A second aspect of flexibility comes from observing that a subset of RMF subcategories only becomes relevant once the AI system has reached a particular development stage. For example, RMF subcategory MANAGE 4.1 is only relevant after the system has been deployed:
\begin{quote}
MANAGE 4.1: Post-deployment AI system monitoring plans are implemented, including mechanisms for capturing and evaluating input from users and other relevant AI actors, appeal and override, decommissioning, incident response, recovery, and change management. \cite{nist_cybersecurity_2023}
\end{quote}

For this reason, the questionnaire is divided into phases of the development lifecycle, based on the AI life cycle described in the RMF \cite{nist_airmf_2023}. We grouped the life cycle into three stages: (1) Planning and design; (2) Data collection and model building, including verifying and validating the system; and (3) Deployment - including deploying, using, operating, and monitoring the system. Each life-cycle stage contains topics and statements appropriate for that stage from multiple RMF pillars. The evaluator only uses the statements suitable for the relevant life-cycle stage. This flexibility explicitly guides evaluators to avoid questions that are not yet relevant to a particular AI system. 

\subsection{Flexibility 3: Multiplicity of AI systems}
Organizations may have multiple AI systems, and the questionnaire allows for flexibility in approaching this multiplicity. Evaluators may score each AI system to be scored separately and aggregate those to get scores for the organization as a whole. Those interested in a more coarse-grain evaluation may instead score the organization holistically without delving into the details of each individual system.

\subsection{Putting it together}

Putting together all three aspects of flexibility, those interested in the most fine-grained version of the evaluation will score each AI system using the individual statements appropriate to that system’s life-cycle stage. Those interested in the most coarse-grained version of the evaluation will score the organization as a whole using only topic statements appropriate to the life-cycle stage of the most advanced AI system that the organization manages. 

\begin{figure}
    \centering
    \includegraphics[width=1\linewidth]{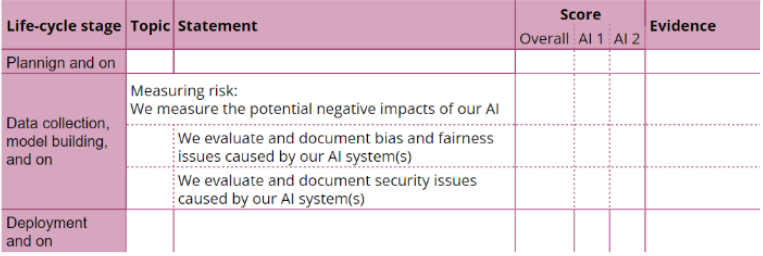}
    \caption{The structure of the questionnaire}
    \label{fig:questionnaire-structure}
\end{figure}

You can see an illustration of the overall structure of the questionnaire in Figure \ref{fig:questionnaire-structure}.

\section{Scoring Guidelines}

\subsection{Scoring Metrics}
In addition to a questionnaire, the maturity model also includes scoring guidelines. Scoring AI responsibility involves a great degree of personal judgment. The goal is not to produce ``objective'' scores but rather to communicate evaluations and reasons for those evaluations that can help organizations understand where they are and how they can improve. Our scoring guidelines are designed to facilitate this kind of evaluation by guiding evaluators to provide evidence-based, well-reasoned evaluations that are based on expectations drawn from NIST's work. 

When scoring, each evaluated statement should be ranked on a scale of 1-5, where 1 is the lowest and 5 is the highest, based on how well it satisfies three metrics (explained in more detail below): coverage of the RMF categories, robustness based on NIST's implementation tiers, and input diversity. For each of these metrics, evaluators should determine the degree to which it is satisfied -- low, medium, or high. We explain how to put it all together below, after presenting the three metrics. We illustrate using these metrics on a case study in appendix \ref{Scoring-examples-appendix}. 

\subsubsection{Coverage of RMF Subcategories}

As discussed above, the questionnaire allows evaluators to evaluate topic statements only, rather than all of the individual statements. For example, the evaluator can score the statement ''\emph{Measuring Risk}: we measure the potential negative impacts of our AI”, but not all the statements it contains, such as ''We evaluate and document bias and fairness issues caused by our AI system(s)” and ''We evaluate and document security issues caused by our AI system(s).” When the evaluator does so, the scoring of the topic statement should reflect coverage of all the individual statements included in that topic. For example, companies that evaluate and document security but not fairness risks satisfy this metric to a degree lower than companies that address both.

\subsubsection{Robustness}
We use the name “robustness” to refer to the ideals expressed through NIST’s ''implementation tiers.” The implementation tiers are distinctions NIST uses to describe degrees of risk management activities in areas such as privacy and cyber-security (\cite{nist_privacy_2020}, \cite{nist_cybersecurity_2023}). These tiers represent an increasing degree of rigor, and showcases how well an organization has implemented the component under evaluation. There are four tiers:
\begin{enumerate}
    \item  PARTIAL- Activities are ad-hoc, reactive, occasional, or isolated from key organizational activities. For example, in organizations with a ''partial” level of privacy risk management, ''[o]rganizational privacy risk management practices are not formalized, and risk is managed in an ad hoc and sometimes reactive manner.” \cite{nist_privacy_2020}
    \item  RISK INFORMED - Activities occur but they are informal and irregular. For example, in organizations with an ''informed” level of privacy risk management, ''[p]rivacy risk assessment occurs, but is not typically repeatable or reoccurring.” \cite{nist_privacy_2020}
    \item  REPEATABLE- Activities are formalized into organization-wide policies or systematic practices. For example, in organizations with a ''repeatable” level of cybersecurity risk management ''[r]isk-informed policies, processes, and procedures are defined, implemented as intended, and reviewed.” \cite{nist_cybersecurity_2023}
    \item  ADAPTIVE - Risk management activities can adapt to changes in the landscape and product, including regular reviews and effective contingency processes to respond to failure. For example, in organizations with an ''adaptive” level of cybersecurity risk  ''The organization uses real-time or near real-time information to understand and consistently act upon cybersecurity risks associated with the products and services it provides and uses.” \cite{nist_cybersecurity_2023} 
\end{enumerate}

The implementation tiers are meant to be tools for internal communication to help organizations set priorities. Organizations are not expected to treat these tiers as targets, but should evaluate the desired tier that is appropriate based on their organizational goals, to both reduce risk to an acceptable level and that is feasible to implement and manage. However, the ideals they express can be used to evaluate maturity: The more an organization embodies the ideals, the more mature it is. We have extracted six interrelated ideals for the purposes of maturity evaluation. For convenience, we refer to them collectively as “robustness”:

\hfill 

\emph{Robustness} - The risk management activities are

\begin{enumerate}
    \item Regular - Performed in a routine manner
    \item Systematic - Follow policies that are well-defined and span company-wide
    \item Trained Personnel - Performed by people who are properly trained and whose roles in the activities are clearly defined
    \item Sufficiently Resourced - Supported by sufficient resources, including budget, time, compute power, and cutting-edge tools
    \item Adaptive - Adapting to changes in the landscape and product, including regular reviews and effective contingency processes to respond to failure
    \item Cross-functional - Involve all core business units and senior management. They are informed of the outcomes and contribute to decision-making, strategy, and resource allocation related to the activities (core business units include finance, customer support, HR, marketing, sales, etc)
\end{enumerate}

\subsubsection{Input diversity}
Input diversity means that risk management activities receive input from diverse internal and external stakeholders. A low level of input diversity means that the relevant activities receive input from relatively few kinds of stakeholders. High levels of input diversity mean that the activities receive input from diverse internal and external stakeholders. For example, suppose that a company chooses its fairness metrics in consultation with civil society organizations, surveys of diverse customers administered by the customer success team, and conversations with diverse employees in the company. In that case, the company demonstrates a high level of input diversity with regard to the statement “We evaluate and document bias and fairness issues related to this AI system”.

The input diversity ideal is not highlighted in the NIST implementation tiers for privacy and cybersecurity. However, it is a key aspect of the AI RMF and is important in AI ethics.  AI systems often impact masses of end-users and data subjects as well as society at large. Properly understanding, measuring, and managing AI risks requires an in-depth understanding of the potential impacts which, in turn, requires input from a wide range of perspectives. 

One way to add input diversity to the maturity model is to include statements that specifically target activities that solicit feedback and incorporate it into the design of the AI system. There are even two RMF subcategories that articulate this content (GOVERN 5.1 and 5.2). However, we choose to include input diversity as a scoring metric because it is a topic that intersects most, or even all, the subcategories. Including input diversity as a scoring guideline allows the evaluator to reflect, and look for evidence for, how well the company solicits and uses feedback on its AI systems and the risks associated with them.  

\subsection{Scores and Evidence}

The score of each statement depends on how well the three metrics are satisfied, and evaluators are asked to provide the evidence and rationale for the scoring. 

Scores ranges between 1-5, where 1 is the lowest and 5 is the highest. We developed the following as a rule of thumb for determining scores:

\begin{itemize}
    \item 5: HHH
    \subitem All three metrics are satisfied to a high degree
    \item 4: HHM
    \subitem Two of the metrics are  satisfied to a high degree and one to a medium degree
    \item 3: HMM, HHL, HML, or MMM
    \subitem One of the following is the case: (1) Two of the metrics are satisfied to a medium degree and one to a high degree; (2) Two of the metrics are satisfied to a high degree and one to a low degree; (3) One metric is satisfied to a high degree, one to a medium degree, and one to a low degree; or (4) all metrics are satisfied to a medium degree.
    \item 2: MML, MLL, or HLL
    \subitem One of the following is the case: (1) Two of the metrics are satisfied to a medium degree and one to a low degree; (2) One metric is satisfied to a medium degree and two to a low degree; (3) One of the metrics is satisfied to a high degree and two to a low degree.
    \item 1: LLL
    \subitem All metrics are satisfied to a low degree
    \item N/A
    \subitem The statement is not applicable
\end{itemize}

This rule of thumb is based on thresholds which we determined in the following way. The satisfaction of one metric to a low degree counts as one point, medium counts as two points, and high counts as three points. Score 1 signifies a total of 3 points, Score 2 signifies 4-5 points, score 3 signifies 6-7 points, score 4 signifies 8 points, and score 5 signifies 9 points. For simplicity, we do not ask evaluators to keep track of these points. Instead, they are asked to use the list above.

 Evidence includes information about what organizations do, about what they don't do, and reports of lack of evidence. For example, evidence may include describing artifacts that indicate that the company is engaged in the relevant activities or the evaluator’s first-hand experience in the company. E.g., they may describe which company documents contain the relevant information and how detailed that information is, the evaluator's first-hand knowledge about the execution of the relevant tasks, and so on. Evidence may also include indications that certain activities are not performed, which may happen, for example, when company documents imply that these activities are outside of the company’s current scope. Further, evidence discussions may also include pointing out a lack of evidence. We ask evaluators to note in their comments a distinction between lack of any evidence and presence of evidence to the contrary. Last, evaluators provide evidence that a statement is not applicable, which may happen for example, due to the life-cycle stage of the evaluated AI system.

We chose to base the scoring on evidence and ask the evaluators to describe or provide it for accountability and to increase the usefulness of the evaluation. Providing evidence encourages accountability in the evaluation process because it requires the evaluator to base the scoring on information that others can assess, too. Moreover, requiring evaluators to provide evidence also encourages accountability on the part of the evaluated companies, because it encourages them to ensure that such evidence is available. Companies can do so, for example, by documenting key processes and their outcomes. 

Providing evidence for scoring improves the usefulness of the evaluation because it contextualizes and explains the reason for the score. Numbers on their own don’t offer much information about the company, what they currently do, what is missing, and how they can improve. The evidence an evaluator cites helps others understand how the evaluator interprets the scoring guidelines and what a given score means to that evaluator. This can help companies understand what they are doing right and how to do better.

\subsection{Applicability of the Scoring Guidelines}

Inevitably, there is going to be some divergence in the scores when completed by different evaluators. This will be true for any set of guidelines, as no set of guidelines can cover all the details relevant to the wide range of contexts and circumstances evaluators may encounter. No matter how detailed the guidelines may be, evaluators will always need to exercise some judgment, deciding what satisfies the metrics and to what degree, deciding what counts as evidence, deciding which contextual factors matter most, etc. These are some of the reasons why scoring AI responsibility is an inherently subjective activity. 

This subjectivity is a feature, not a bug, and the goal of the scoring guidelines is to help evaluators express these subjective judgments in a structured and helpful way. To further support the process, we articulate our own judgment about two examples in appendix \ref{Scoring-examples-appendix}. In the last section, Section \ref{limitations-section}, we discuss ongoing and future work to provide further information for evaluators.

\section{Score Aggregation}

\begin{figure}
    \centering
    \includegraphics[width=1\linewidth]{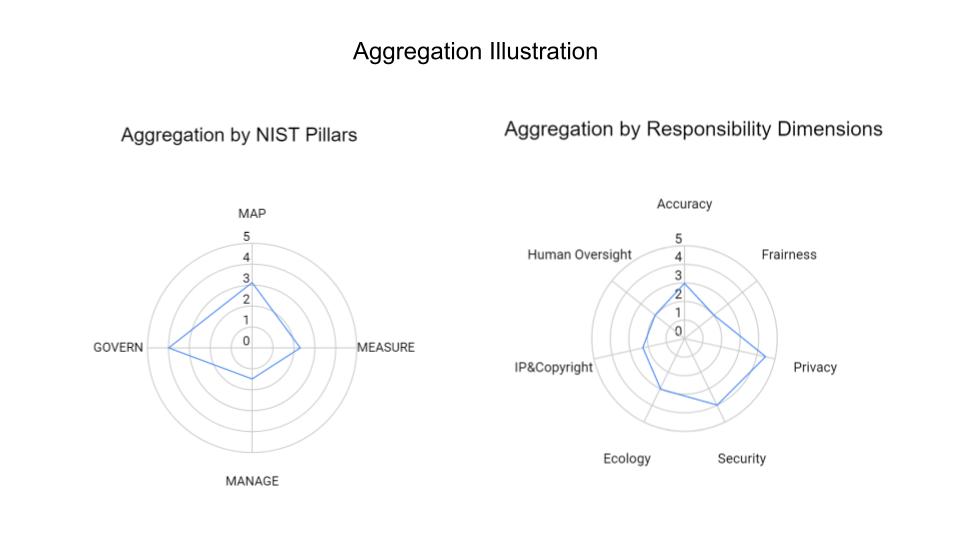}
    \caption{Illustration of aggregation modes in radio charts: To the left, aggregation by NIST Pillar. To the right, aggregation by responsibility dimension}
    \label{fig:aggregation}
\end{figure}

After scoring the individual statements or topics, evaluators can aggregate the scoring to get a unified score. Our maturity model offers two modes of aggregation: By NIST pillars and by responsibility dimensions (see Figure \ref{fig:aggregation} for an illustration).

In aggregating by NIST pillars, the output is a score for each of the NIST pillars, MAP, MEASURE, MANAGE, and GOVERN, based on the scores of the statements that belong to it. For example, the MAP score is the average of all the statements that are based on recommendations included in the MAP pillar. 

Aggregation by NIST pillar can help discover organizations’ strengths and weaknesses in different kinds of activities. In particular, this mode of aggregation can expose systematic failures in organizations’ approaches to AI responsibility. For example, when organizations show strength in GOVERN activities but weakness in all other pillars, they may be engaged in ethics washing. For example, they may be establishing policies that are largely not implemented. Other organizations may show strength in GOVERN and MANAGE but weakness in MAP and MEASURE. These organizations’ risk management activities may be ill-informed, as the low level of MAP and MEASURE may indicate that their understanding of the risks is lacking.

Another option is to aggregate based on some or all the dimensions of AI responsibility the RMF identifies, e.g., fairness, privacy, and security. In this aggregation mode, the score of each dimension is an average of all the statements relevant to that dimension and is possible only when the relevant individual statements get their own score. 

Aggregation by responsibility dimensions can help discover when organizations ignore certain issues. For example, some organizations boast AI responsibility based on their activity in a handful of risk areas, such as privacy and security. Focus on each dimension can highlight the other areas, in which the company is lacking.

\begin{figure}
    \centering
    \includegraphics[width=1\linewidth]{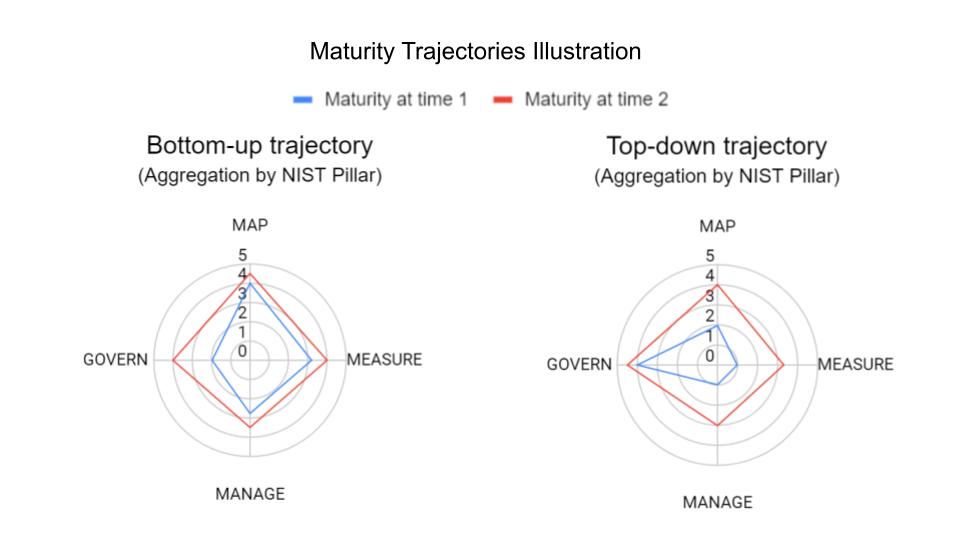}
    \caption{Illustration of maturity progress trajectories. To the left, a bottom-up trajectory. To the right, a top-town trajectory}
    \label{fig:trajectories}
\end{figure}

Aggregation can help track companies’ progress over time. Our maturity model isn't prescriptive about the trajectory of the progress. It allows tracking progress which may take place in many different ways. For example, In large corporations, for example, we may see a top-down progress trajectory, where the company starts with strong GOVERN activities and advances to stronger MEASURE and MANAGE activities. In smaller companies, we might see a bottom-up progress trajectory, where the company starts with strong MEASURE, MAP, and MANAGE activities and progresses to stronger GOVERN activities (see Figure \ref{fig:trajectories} for an illustration). 

\section{Advantages, limitations, and future work}
\label{limitations-section}

This paper lays the foundation for a maturity model for responsible AI governance. It is based on the NIST AI RMF, a widely accepted voluntary framework, which increases its ability to facilitate comparability between organizations, gets away from questions of legal compliance, and highlights the mitigation of sociotechnical harms. Further, the model is flexible in that evaluators can adjust the questionnaire to the context and in that the evaluation accommodates multiple maturity trajectories. 

We see this paper as laying the groundwork for future, practice-informed iterations of the maturity model. A key area for future work is the scoring guidelines. In their current form, the scoring guidelines are likely to lead to vast variations in evaluator behaviors, e.g., evaluators may interpret what counts as “evidence” differently or they might treat the same evidence as supporting different levels of satisfaction of the metrics. We anticipate that evaluators might face uncertainty in how to best apply the current guidelines. Moreover, we anticipate uncertainty in interpreting aggregate scores. Organizations might not know what to make of having a certain aggregated score. 

Due to the inevitable subjectivity of evaluations, some degree of uncertainty will always remain. Having said that, in ongoing work based on the foundation laid out in this paper, we seek to better understand the perspective of people who aren’t us about what it’s like to use this model, and especially scoring. In that work, we are facilitating group experimentation with the model: We are recording how evaluators use it and their impressions of it, including what they count as “evidence” and what merits satisfying the metrics to different degrees. Our analysis will reveal trends in scoring and evidence-giving that will help set expectations for scoring, such as which activities typically merit various scores and which kinds of evidence generally are given to support them. This practice will allow us to embed more diverse perspectives into the model and the scoring guidelines in particular. Moreover, in future work, we will use the model to evaluate a wide range of companies. This will help create scoring expectations by different company characteristics such as size and sector. 

As with other evaluative frameworks, our maturity model carries the risk of ethics washing. Our maturity model is strongly process-oriented. The model doesn’t evaluate companies on the success of their risk mitigation efforts, e.g. how ``fair'' their model is. This choice is intentional. First, we follow the NIST AI RMF which makes the same choice. Second, concepts related to socio-technical harm, such as fairness, are deeply context-sensitive. Allowing organizations to determine for themselves how to interpret and measure key concepts is crucial for pluralism. The downside of this choice is that it creates opportunities for ethics washing. Per Goodhart’s law \cite{goodhart_problems_1984, hennessy_goodharts_2020, chrystal_goodharts_2003}, ``[w]hen a measure becomes a target, it ceases to be a good measure'' \cite{strathern_improving_1997}. Our framework is no exception. Organizations may find ways to appear to have high responsible AI maturity in accordance with the model, while still having poor performance in terms of their sociotechnical harm. However, the same is true for metrics that measure harms directly. Our model decreases gaming risks by moving away from strict compliance by using a voluntary framework as a foundation, which decreases compliance pressures. 

\section{Conclusion}
This paper lays out a foundation for a maturity model to evaluate the responsibility of AI governance in organizations that develop and manage AI systems. This foundation includes a flexible questionnaire and scoring guidelines, both based on industry standards set out by NIST. The strengths of this model include a rigorous conceptual framework that is drawn from industry standards, a focus on the mitigation of sociotechnical harm and inclusivity, flexibility in the questionnaire and aggregation options to accommodate the needs of different organizations, compatibility with multiple maturity trajectories, and the facilitation of evidence-based evaluations that flesh out subjective judgments and the reasoning to support them. All these are intended to make this model practical and helpful in the hopes of supporting organizations in improving their AI risk management and supporting the field in enhancing the overall levels of AI ethics implementation, which are currently dangerously low.

\section{Research Ethics and Social Impact}

\subsection{Ethical Considerations}
Our work involved consultation and feedback from practitioners. Therefore, we have consulted with an Institutional Review Board, and they have determined that this project is not Human Subject Research. To ensure transparency, we have conveyed to all our interlocutors that the conversation is being held in the context of a research paper and that insights based on the conversation may appear in the paper. We have obtained consent before proceeding with the conversation.  


\subsection{Adverse Impacts}
As we discussed in the body of the paper, when used inappropriately, the framework we developed in this paper could be used to provide a false assurance of compliance. The framework should be used to guide processes of improvement of AI governance. The explanations and evidence evaluators are asked to provide are aimed to facilitate that. Usages that ignore that, such as treating the framework as a checklist, are misguided and could especially lead to the misrepresentation of the evaluated company and lack of progress.

\section{Acknowledgments}
Ravit Dotan was supported by a grant from the Notre Dame-IBM Tech Ethics Lab. 

We are thankful to our collaborators on applied aspects of this ongoing project for extensive feedback on this paper: Carol Anderson, Benny Esparra, and Ric McLaughlin. We also thank the following individuals and organizations for their help and feedback: (in alphabetical order) Anita Dorett, Jesse Dunietz, Diana Glassman, Saurabh Gupta, KOKO Labs (Palo Alto, CA), Navishka Pandit, Larisa Ruoff,  Reva Schwartz.


    


\bibliographystyle{ACM-Reference-Format}
\bibliography{bibliography}


\begin{thebibliography}{48}


\ifx \showCODEN    \undefined \def \showCODEN     #1{\unskip}     \fi
\ifx \showDOI      \undefined \def \showDOI       #1{#1}\fi
\ifx \showISBNx    \undefined \def \showISBNx     #1{\unskip}     \fi
\ifx \showISBNxiii \undefined \def \showISBNxiii  #1{\unskip}     \fi
\ifx \showISSN     \undefined \def \showISSN      #1{\unskip}     \fi
\ifx \showLCCN     \undefined \def \showLCCN      #1{\unskip}     \fi
\ifx \shownote     \undefined \def \shownote      #1{#1}          \fi
\ifx \showarticletitle \undefined \def \showarticletitle #1{#1}   \fi
\ifx \showURL      \undefined \def \showURL       {\relax}        \fi
\providecommand\bibfield[2]{#2}
\providecommand\bibinfo[2]{#2}
\providecommand\natexlab[1]{#1}
\providecommand\showeprint[2][]{arXiv:#2}

\bibitem[CMM({[n.\,d.]})]%
        {CMMI_cmmi_nodate}
 \bibinfo{year}{[n.\,d.]}\natexlab{}.
\newblock \bibinfo{title}{{CMMI} {Institute} - {Home}}.
\newblock
\newblock
\urldef\tempurl%
\url{https://cmmiinstitute.com/}
\showURL{%
\tempurl}


\bibitem[Ayling and Chapman(2022)]%
        {ayling_putting_2022}
\bibfield{author}{\bibinfo{person}{Jacqui Ayling} {and} \bibinfo{person}{Adriane Chapman}.} \bibinfo{year}{2022}\natexlab{}.
\newblock \showarticletitle{Putting {AI} ethics to work: are the tools fit for purpose?}
\newblock \bibinfo{journal}{\emph{AI and Ethics}} \bibinfo{volume}{2}, \bibinfo{number}{3} (\bibinfo{date}{Aug.} \bibinfo{year}{2022}), \bibinfo{pages}{405--429}.
\newblock
\showISSN{2730-5953, 2730-5961}
\urldef\tempurl%
\url{https://doi.org/10.1007/s43681-021-00084-x}
\showDOI{\tempurl}


\bibitem[{Baxter, K}(2021)]%
        {baxter_k_ai_2021}
\bibfield{author}{\bibinfo{person}{{Baxter, K}}.} \bibinfo{year}{2021}\natexlab{}.
\newblock \bibinfo{booktitle}{\emph{{AI} {Ethics} {Maturity} {Model}}}.
\newblock \bibinfo{type}{{T}echnical {R}eport}. \bibinfo{institution}{Salesforce}.
\newblock
\urldef\tempurl%
\url{https://www.salesforceairesearch.com/static/ethics/EthicalAIMaturityModel.pdf}
\showURL{%
\tempurl}


\bibitem[Bommasani(2022)]%
        {bommasani_evaluation_2022}
\bibfield{author}{\bibinfo{person}{Rishi Bommasani}.} \bibinfo{year}{2022}\natexlab{}.
\newblock \showarticletitle{Evaluation for {Change}}.
\newblock  (\bibinfo{year}{2022}).
\newblock
\urldef\tempurl%
\url{https://doi.org/10.48550/ARXIV.2212.11670}
\showDOI{\tempurl}
\newblock
\shownote{Publisher: arXiv Version Number: 1}.


\bibitem[Bos-Brouwers(2009)]%
        {Bos-Brouwers}
\bibfield{author}{\bibinfo{person}{Hilke Elke~Jacke Bos-Brouwers}.} \bibinfo{year}{2009}\natexlab{}.
\newblock \showarticletitle{Corporate sustainability and innovation in SMEs: Evidence of themes and activities in practice}.
\newblock \bibinfo{journal}{\emph{Business Strategy and the Environment}}  \bibinfo{volume}{19} (\bibinfo{date}{June} \bibinfo{year}{2009}), \bibinfo{pages}{417--435}.
\newblock
\showISSN{7}
\urldef\tempurl%
\url{https://doi.org/10.1002/bse.652}
\showDOI{\tempurl}


\bibitem[{Burstein, Jill}(2023)]%
        {burstein_jill_duolingo_2023}
\bibfield{author}{\bibinfo{person}{{Burstein, Jill}}.} \bibinfo{year}{2023}\natexlab{}.
\newblock \bibinfo{booktitle}{\emph{Duolingo {English} {Test} {Responsible} {AI} {Standards}}}.
\newblock \bibinfo{type}{{T}echnical {R}eport}. \bibinfo{institution}{Duolingo}.
\newblock
\urldef\tempurl%
\url{https://duolingo-papers.s3.amazonaws.com/other/DET+Responsible+AI+033123.pdf}
\showURL{%
\tempurl}


\bibitem[{CCMC}(2021)]%
        {ccmc_cmmc_2021}
\bibfield{author}{\bibinfo{person}{{CCMC}}.} \bibinfo{year}{2021}\natexlab{}.
\newblock \bibinfo{booktitle}{\emph{{CMMC} {PROGRAM} {PROPOSED} {RULE} {PUBLISHED} - {PUBLIC} {COMMENT} {PERIOD} {BEGINS}}}.
\newblock \bibinfo{type}{{T}echnical {R}eport}. \bibinfo{institution}{US Department of Defense}.
\newblock
\urldef\tempurl%
\url{https://dodcio.defense.gov/CMMC/about/}
\showURL{%
\tempurl}


\bibitem[Chrystal and Mizen(2003)]%
        {chrystal_goodharts_2003}
\bibfield{author}{\bibinfo{person}{Alec Chrystal} {and} \bibinfo{person}{Paul Mizen}.} \bibinfo{year}{2003}\natexlab{}.
\newblock \showarticletitle{Goodhart's {Law}: its origins, meaning and implications for monetary policy}.
\newblock In \bibinfo{booktitle}{\emph{Central {Banking}, {Monetary} {Theory} and {Practice}}}. \bibinfo{publisher}{Edward Elgar Publishing}, \bibinfo{pages}{2329}.
\newblock
\showISBNx{978-1-78195-077-7}
\urldef\tempurl%
\url{https://doi.org/10.4337/9781781950777.00022}
\showDOI{\tempurl}


\bibitem[Covello and Iatridis(2020)]%
        {Covello&Iatridis}
\bibfield{author}{\bibinfo{person}{Christina Covello} {and} \bibinfo{person}{Konstantinos Iatridis}.} \bibinfo{year}{2020}\natexlab{}.
\newblock \showarticletitle{On the challenges and drivers of implementing responsible innovation in foodpreneurial SMEs}.
\newblock In \bibinfo{booktitle}{\emph{Assessment of Responsible Innovation} (\bibinfo{edition}{1} ed.)}, \bibfield{editor}{\bibinfo{person}{Emad Yaghmaei} {and} \bibinfo{person}{Ibo van~de Poel}} (Eds.). \bibinfo{publisher}{Routledge}.
\newblock
\showISBNx{9780429298998}
\urldef\tempurl%
\url{https://doi.org/10.4324/9780429298998}
\showDOI{\tempurl}


\bibitem[Crosby(1979)]%
        {crosby_quality_1979}
\bibfield{author}{\bibinfo{person}{Philip~B. Crosby}.} \bibinfo{year}{1979}\natexlab{}.
\newblock \bibinfo{booktitle}{\emph{Quality is free: the art of making quality certain}}.
\newblock \bibinfo{publisher}{McGraw-Hill}, \bibinfo{address}{New York}.
\newblock
\showISBNx{978-0-07-014512-2}


\bibitem[Dotan et~al\mbox{.}(2024)]%
        {dotan_evaluating_2024}
\bibfield{author}{\bibinfo{person}{Ravit Dotan}, \bibinfo{person}{Gil Rosenthal}, \bibinfo{person}{Tess Buckley}, \bibinfo{person}{Josh Scarpino}, \bibinfo{person}{Luke Patterson}, {and} \bibinfo{person}{Thorin Bristow}.} \bibinfo{year}{2024}\natexlab{}.
\newblock \bibinfo{booktitle}{\emph{Evaluating {AI} {Governance}: {Insights} from {Public} {Disclosures}}}.
\newblock \bibinfo{type}{{T}echnical {R}eport}.
\newblock
\urldef\tempurl%
\url{https://www.ravitdotan.com/_files/ugd/f83391_b853450bcc274e9ba9454d618ee41a94.pdf}
\showURL{%
\tempurl}


\bibitem[Drew(2018)]%
        {drew_design_2018}
\bibfield{author}{\bibinfo{person}{Cat Drew}.} \bibinfo{year}{2018}\natexlab{}.
\newblock \showarticletitle{Design for data ethics: using service design approaches to operationalize ethical principles on four projects}.
\newblock \bibinfo{journal}{\emph{Philosophical Transactions of the Royal Society A: Mathematical, Physical and Engineering Sciences}} \bibinfo{volume}{376}, \bibinfo{number}{2128} (\bibinfo{date}{Sept.} \bibinfo{year}{2018}), \bibinfo{pages}{20170353}.
\newblock
\showISSN{1364-503X, 1471-2962}
\urldef\tempurl%
\url{https://doi.org/10.1098/rsta.2017.0353}
\showDOI{\tempurl}


\bibitem[{Ethical Intelligence} et~al\mbox{.}(2022)]%
        {ethical_intelligence_ethics_2022}
\bibfield{author}{\bibinfo{person}{{Ethical Intelligence}}, \bibinfo{person}{{BCV}}, {and} \bibinfo{person}{{EAIGG}}.} \bibinfo{year}{2022}\natexlab{}.
\newblock \bibinfo{booktitle}{\emph{{ETHICS} {MATURITY} {CONTINUUM}}}.
\newblock \bibinfo{type}{{T}echnical {R}eport}.
\newblock
\urldef\tempurl%
\url{https://static1.squarespace.com/static/5f6dbf464a8eec79c3d177c0/t/61e8821d53b74041072d556d/1642627614838/Ethics+Maturity+Continuum+Report.pdf}
\showURL{%
\tempurl}


\bibitem[Gebru et~al\mbox{.}(2021)]%
        {gebru_datasheets_2021}
\bibfield{author}{\bibinfo{person}{Timnit Gebru}, \bibinfo{person}{Jamie Morgenstern}, \bibinfo{person}{Briana Vecchione}, \bibinfo{person}{Jennifer~Wortman Vaughan}, \bibinfo{person}{Hanna Wallach}, \bibinfo{person}{Hal~Daumé Iii}, {and} \bibinfo{person}{Kate Crawford}.} \bibinfo{year}{2021}\natexlab{}.
\newblock \showarticletitle{Datasheets for datasets}.
\newblock \bibinfo{journal}{\emph{Commun. ACM}} \bibinfo{volume}{64}, \bibinfo{number}{12} (\bibinfo{date}{Dec.} \bibinfo{year}{2021}), \bibinfo{pages}{86--92}.
\newblock
\showISSN{0001-0782, 1557-7317}
\urldef\tempurl%
\url{https://doi.org/10.1145/3458723}
\showDOI{\tempurl}


\bibitem[Goodhart(1984)]%
        {goodhart_problems_1984}
\bibfield{author}{\bibinfo{person}{C.~A.~E. Goodhart}.} \bibinfo{year}{1984}\natexlab{}.
\newblock \showarticletitle{Problems of {Monetary} {Management}: {The} {UK} {Experience}}.
\newblock In \bibinfo{booktitle}{\emph{Monetary {Theory} and {Practice}}}. \bibinfo{publisher}{Macmillan Education UK}, \bibinfo{address}{London}, \bibinfo{pages}{91--121}.
\newblock
\showISBNx{978-0-333-36060-6 978-1-349-17295-5}
\urldef\tempurl%
\url{https://doi.org/10.1007/978-1-349-17295-5_4}
\showDOI{\tempurl}


\bibitem[Hennessy and Goodhart(2020)]%
        {hennessy_goodharts_2020}
\bibfield{author}{\bibinfo{person}{Christopher Hennessy} {and} \bibinfo{person}{Charles~A.E. Goodhart}.} \bibinfo{year}{2020}\natexlab{}.
\newblock \showarticletitle{Goodhart's {Law} and {Machine} {Learning}}.
\newblock \bibinfo{journal}{\emph{SSRN Electronic Journal}} (\bibinfo{year}{2020}).
\newblock
\showISSN{1556-5068}
\urldef\tempurl%
\url{https://doi.org/10.2139/ssrn.3639508}
\showDOI{\tempurl}


\bibitem[Holland et~al\mbox{.}(2018)]%
        {holland_dataset_2018}
\bibfield{author}{\bibinfo{person}{Sarah Holland}, \bibinfo{person}{Ahmed Hosny}, \bibinfo{person}{Sarah Newman}, \bibinfo{person}{Joshua Joseph}, {and} \bibinfo{person}{Kasia Chmielinski}.} \bibinfo{year}{2018}\natexlab{}.
\newblock \bibinfo{title}{The {Dataset} {Nutrition} {Label}: {A} {Framework} {To} {Drive} {Higher} {Data} {Quality} {Standards}}.
\newblock
\newblock
\urldef\tempurl%
\url{https://doi.org/10.48550/arXiv.1805.03677}
\showDOI{\tempurl}
\newblock
\shownote{arXiv:1805.03677 [cs]}.


\bibitem[{IBM}({[n.\,d.]})]%
        {ibm_ibm_2022}
\bibfield{author}{\bibinfo{person}{{IBM}}.} \bibinfo{year}{[n.\,d.]}\natexlab{}.
\newblock \bibinfo{title}{{IBM} {Global} {AI} {Adoption} {Index} 2022 {\textbar} {IBM}}.
\newblock
\newblock
\urldef\tempurl%
\url{https://www.ibm.com/watson/resources/ai-adoption}
\showURL{%
\tempurl}


\bibitem[IBM(2021)]%
        {ibm_ai_2021}
\bibfield{author}{\bibinfo{person}{IBM}.} \bibinfo{year}{2021}\natexlab{}.
\newblock \bibinfo{booktitle}{\emph{{AI} maturity framework for enterprise applications}}.
\newblock \bibinfo{type}{{T}echnical {R}eport}. \bibinfo{institution}{IBM}.
\newblock
\urldef\tempurl%
\url{https://www.ibm.com/watson/supply-chain/resources/ai-maturity}
\showURL{%
\tempurl}


\bibitem[Krijger et~al\mbox{.}(2023)]%
        {krijger_ai_2023}
\bibfield{author}{\bibinfo{person}{J. Krijger}, \bibinfo{person}{T. Thuis}, \bibinfo{person}{M. de Ruiter}, \bibinfo{person}{E. Ligthart}, {and} \bibinfo{person}{I. Broekman}.} \bibinfo{year}{2023}\natexlab{}.
\newblock \showarticletitle{The {AI} ethics maturity model: a holistic approach to advancing ethical data science in organizations}.
\newblock \bibinfo{journal}{\emph{AI and Ethics}} \bibinfo{volume}{3}, \bibinfo{number}{2} (\bibinfo{date}{May} \bibinfo{year}{2023}), \bibinfo{pages}{355--367}.
\newblock
\showISSN{2730-5961}
\urldef\tempurl%
\url{https://doi.org/10.1007/s43681-022-00228-7}
\showDOI{\tempurl}


\bibitem[Kuznets(1965)]%
        {kuznets_economic_1965}
\bibfield{author}{\bibinfo{person}{Simon~Smith Kuznets}.} \bibinfo{year}{1965}\natexlab{}.
\newblock \bibinfo{booktitle}{\emph{Economic growth and structure: selected essays}}.
\newblock \bibinfo{publisher}{Norton}, \bibinfo{address}{New York, N.Y}.
\newblock
\showISBNx{978-0-393-09475-6}


\bibitem[Lazar and Nelson(2023)]%
        {lazar_ai_2023}
\bibfield{author}{\bibinfo{person}{Seth Lazar} {and} \bibinfo{person}{Alondra Nelson}.} \bibinfo{year}{2023}\natexlab{}.
\newblock \showarticletitle{{AI} safety on whose terms?}
\newblock \bibinfo{journal}{\emph{Science}} \bibinfo{volume}{381}, \bibinfo{number}{6654} (\bibinfo{date}{July} \bibinfo{year}{2023}), \bibinfo{pages}{138--138}.
\newblock
\showISSN{0036-8075, 1095-9203}
\urldef\tempurl%
\url{https://doi.org/10.1126/science.adi8982}
\showDOI{\tempurl}


\bibitem[{McKinsey}({[n.\,d.]})]%
        {mckinsey_state_nodate}
\bibfield{author}{\bibinfo{person}{{McKinsey}}.} \bibinfo{year}{[n.\,d.]}\natexlab{}.
\newblock \bibinfo{title}{The state of {AI} in 2022—and a half decade in review {\textbar} {McKinsey}}.
\newblock
\newblock
\urldef\tempurl%
\url{https://www.mckinsey.com/capabilities/quantumblack/our-insights/the-state-of-ai-in-2022-and-a-half-decade-in-review#/download//~/media/mckinsey/business%20functions/quantumblack/our%20insights/the%20state%20of%20ai%20in%202022%20and%20a%20half%20decade%20in%20review/the-state-of-ai-in-2022-and-a-half-decade-in-review.pdf?cid=soc-web}
\showURL{%
\tempurl}


\bibitem[Mitchell et~al\mbox{.}(2019)]%
        {mitchell_model_2019}
\bibfield{author}{\bibinfo{person}{Margaret Mitchell}, \bibinfo{person}{Simone Wu}, \bibinfo{person}{Andrew Zaldivar}, \bibinfo{person}{Parker Barnes}, \bibinfo{person}{Lucy Vasserman}, \bibinfo{person}{Ben Hutchinson}, \bibinfo{person}{Elena Spitzer}, \bibinfo{person}{Inioluwa~Deborah Raji}, {and} \bibinfo{person}{Timnit Gebru}.} \bibinfo{year}{2019}\natexlab{}.
\newblock \showarticletitle{Model {Cards} for {Model} {Reporting}}. In \bibinfo{booktitle}{\emph{Proceedings of the {Conference} on {Fairness}, {Accountability}, and {Transparency}}}. \bibinfo{pages}{220--229}.
\newblock
\urldef\tempurl%
\url{https://doi.org/10.1145/3287560.3287596}
\showDOI{\tempurl}
\newblock
\shownote{arXiv:1810.03993 [cs]}.


\bibitem[{MITRE}(2023)]%
        {mitre_mitre_2023}
\bibfield{author}{\bibinfo{person}{{MITRE}}.} \bibinfo{year}{2023}\natexlab{}.
\newblock \bibinfo{booktitle}{\emph{The {MITRE} {AI} {Maturity} {Model} and {Organizational} {Assessment} {Tool} {Guide}: {A} {Path} to {Successful} {AI} {Adoption}}}.
\newblock \bibinfo{type}{{T}echnical {R}eport}. \bibinfo{institution}{MITRE}.
\newblock


\bibitem[Morley et~al\mbox{.}(2020)]%
        {morley_what_2020}
\bibfield{author}{\bibinfo{person}{Jessica Morley}, \bibinfo{person}{Luciano Floridi}, \bibinfo{person}{Libby Kinsey}, {and} \bibinfo{person}{Anat Elhalal}.} \bibinfo{year}{2020}\natexlab{}.
\newblock \showarticletitle{From {What} to {How}: {An} {Initial} {Review} of {Publicly} {Available} {AI} {Ethics} {Tools}, {Methods} and {Research} to {Translate} {Principles} into {Practices}}.
\newblock \bibinfo{journal}{\emph{Science and Engineering Ethics}} \bibinfo{volume}{26}, \bibinfo{number}{4} (\bibinfo{date}{Aug.} \bibinfo{year}{2020}), \bibinfo{pages}{2141--2168}.
\newblock
\showISSN{1471-5546}
\urldef\tempurl%
\url{https://doi.org/10.1007/s11948-019-00165-5}
\showDOI{\tempurl}


\bibitem[Morley et~al\mbox{.}(2023)]%
        {morley_operationalising_2023}
\bibfield{author}{\bibinfo{person}{Jessica Morley}, \bibinfo{person}{Libby Kinsey}, \bibinfo{person}{Anat Elhalal}, \bibinfo{person}{Francesca Garcia}, \bibinfo{person}{Marta Ziosi}, {and} \bibinfo{person}{Luciano Floridi}.} \bibinfo{year}{2023}\natexlab{}.
\newblock \showarticletitle{Operationalising {AI} ethics: barriers, enablers and next steps}.
\newblock \bibinfo{journal}{\emph{AI \& SOCIETY}} \bibinfo{volume}{38}, \bibinfo{number}{1} (\bibinfo{date}{Feb.} \bibinfo{year}{2023}), \bibinfo{pages}{411--423}.
\newblock
\showISSN{0951-5666, 1435-5655}
\urldef\tempurl%
\url{https://doi.org/10.1007/s00146-021-01308-8}
\showDOI{\tempurl}


\bibitem[Munn(2023)]%
        {munn_uselessness_2023}
\bibfield{author}{\bibinfo{person}{Luke Munn}.} \bibinfo{year}{2023}\natexlab{}.
\newblock \showarticletitle{The uselessness of {AI} ethics}.
\newblock \bibinfo{journal}{\emph{AI and Ethics}} \bibinfo{volume}{3}, \bibinfo{number}{3} (\bibinfo{date}{Aug.} \bibinfo{year}{2023}), \bibinfo{pages}{869--877}.
\newblock
\showISSN{2730-5961}
\urldef\tempurl%
\url{https://doi.org/10.1007/s43681-022-00209-w}
\showDOI{\tempurl}


\bibitem[{National Institute of Standards and Technology}(2020)]%
        {nist_privacy_2020}
\bibfield{author}{\bibinfo{person}{{National Institute of Standards and Technology}}.} \bibinfo{year}{2020}\natexlab{}.
\newblock \bibinfo{booktitle}{\emph{{NIST} {PRIVACY} {FRAMEWORK}:: {A} {TOOL} {FOR} {IMPROVING} {PRIVACY} {THROUGH} {ENTERPRISE} {RISK} {MANAGEMENT}, {VERSION} 1.0}}.
\newblock \bibinfo{type}{{T}echnical {R}eport} NIST CSWP 01162020. \bibinfo{institution}{National Institute of Standards and Technology}, \bibinfo{address}{Gaithersburg, MD}. \bibinfo{pages}{NIST CSWP 01162020} pages.
\newblock
\urldef\tempurl%
\url{https://doi.org/10.6028/NIST.CSWP.01162020}
\showDOI{\tempurl}


\bibitem[NIST(2023)]%
        {nist_airmf_2023}
\bibfield{author}{\bibinfo{person}{NIST}.} \bibinfo{year}{2023}\natexlab{}.
\newblock \bibinfo{booktitle}{\emph{{AI} {Risk} {Management} {Framework}: {AI} {RMF} (1.0)}}.
\newblock \bibinfo{type}{{T}echnical {R}eport} NIST AI 100-1. \bibinfo{institution}{National Institute of Standards and Technology}, \bibinfo{address}{Gaithersburg, MD}. \bibinfo{pages}{error: NIST AI 100--1} pages.
\newblock
\urldef\tempurl%
\url{https://doi.org/10.6028/NIST.AI.100-1}
\showDOI{\tempurl}


\bibitem[{NIST} et~al\mbox{.}(2023)]%
        {nist_cybersecurity_2023}
\bibfield{author}{\bibinfo{person}{{NIST}}, \bibinfo{person}{{NCCoE}}, \bibinfo{person}{{DOE}}, {and} \bibinfo{person}{{CESER}}.} \bibinfo{year}{2023}\natexlab{}.
\newblock \bibinfo{title}{Cybersecurity {Capability} {Maturity} {Model} to {NIST} {Cybersecurity} {Framework} {Mapping} {\textbar} {NCCoE}}.
\newblock
\newblock
\urldef\tempurl%
\url{https://www.nccoe.nist.gov/news-insights/cybersecurity-capability-maturity-model-nist-cybersecurity-framework-mapping}
\showURL{%
\tempurl}


\bibitem[{Open Data Institute}(2022)]%
        {open_data_institute_data_2022}
\bibfield{author}{\bibinfo{person}{{Open Data Institute}}.} \bibinfo{year}{2022}\natexlab{}.
\newblock \bibinfo{booktitle}{\emph{Data {Ethics} {Maturity} {Model}: benchmarking your approach to data ethics}}.
\newblock \bibinfo{type}{{T}echnical {R}eport}.
\newblock
\urldef\tempurl%
\url{https://theodi.org/insights/tools/data-ethics-maturity-model-benchmarking-your-approach-to-data-ethics/}
\showURL{%
\tempurl}


\bibitem[{OpenAI}({[n.\,d.]})]%
        {openai_preparedness_nodate}
\bibfield{author}{\bibinfo{person}{{OpenAI}}.} \bibinfo{year}{[n.\,d.]}\natexlab{}.
\newblock \bibinfo{title}{Preparedness}.
\newblock
\newblock
\urldef\tempurl%
\url{https://openai.com/safety/preparedness}
\showURL{%
\tempurl}


\bibitem[{OpenAI}(2023)]%
        {openai_our_2023}
\bibfield{author}{\bibinfo{person}{{OpenAI}}.} \bibinfo{year}{2023}\natexlab{}.
\newblock \bibinfo{title}{Our approach to {AI} safety}.
\newblock
\newblock
\urldef\tempurl%
\url{https://openai.com/blog/our-approach-to-ai-safety}
\showURL{%
\tempurl}


\bibitem[Oprysko(2023)]%
        {Oprysko}
\bibfield{author}{\bibinfo{person}{Caitlin Oprysko}.} \bibinfo{year}{2023}\natexlab{}.
\newblock \showarticletitle{OpenAI registers to lobby}.
\newblock \bibinfo{journal}{\emph{Politico}} (\bibinfo{date}{Nov.} \bibinfo{year}{2023}).
\newblock
\urldef\tempurl%
\url{https://www.politico.com/newsletters/politico-influence/2023/11/17/openai-registers-to-lobby-00127874}
\showURL{%
\tempurl}


\bibitem[Piaget(1964)]%
        {piaget_cognitive_1964}
\bibfield{author}{\bibinfo{person}{Jean Piaget}.} \bibinfo{year}{1964}\natexlab{}.
\newblock \showarticletitle{Cognitive development in children: {Piaget} development and learning}.
\newblock \bibinfo{journal}{\emph{Journal of Research in Science Teaching}} \bibinfo{volume}{2}, \bibinfo{number}{3} (\bibinfo{date}{Sept.} \bibinfo{year}{1964}), \bibinfo{pages}{176--186}.
\newblock
\showISSN{0022-4308, 1098-2736}
\urldef\tempurl%
\url{https://doi.org/10.1002/tea.3660020306}
\showDOI{\tempurl}


\bibitem[Proenca et~al\mbox{.}(2017)]%
        {proenca_risk_2017}
\bibfield{author}{\bibinfo{person}{Diogo Proenca}, \bibinfo{person}{Joao Estevens}, \bibinfo{person}{Ricardo Vieira}, {and} \bibinfo{person}{Jose Borbinha}.} \bibinfo{year}{2017}\natexlab{}.
\newblock \showarticletitle{Risk {Management}: {A} {Maturity} {Model} {Based} on {ISO} 31000}. In \bibinfo{booktitle}{\emph{2017 {IEEE} 19th {Conference} on {Business} {Informatics} ({CBI})}}. \bibinfo{publisher}{IEEE}, \bibinfo{address}{Thessaloniki, Greece}, \bibinfo{pages}{99--108}.
\newblock
\showISBNx{978-1-5386-3035-8}
\urldef\tempurl%
\url{https://doi.org/10.1109/CBI.2017.40}
\showDOI{\tempurl}


\bibitem[{PwC}(2021)]%
        {pwc_responsible_2021}
\bibfield{author}{\bibinfo{person}{{PwC}}.} \bibinfo{year}{2021}\natexlab{}.
\newblock \bibinfo{booktitle}{\emph{Responsible {AI} - {Maturing} from theory to practice}}.
\newblock \bibinfo{type}{{T}echnical {R}eport}. \bibinfo{institution}{PwC}.
\newblock


\bibitem[Pöppelbuß and Röglinger(2011)]%
        {poppelbus_what_2011}
\bibfield{author}{\bibinfo{person}{Jens Pöppelbuß} {and} \bibinfo{person}{Maximilian Röglinger}.} \bibinfo{year}{2011}\natexlab{}.
\newblock \showarticletitle{{WHAT} {MAKES} {A} {USEFUL} {MATURITY} {MODEL}? {A} {FRAMEWORK} {OF} {GENERAL} {DESIGN} {PRINCIPLES} {FOR} {MATURITY} {MODELS} {AND} {ITS} {DEMONSTRATION} {IN} {BUSINESS} {PROCESS} {MANAGEMENT}}.
\newblock \bibinfo{journal}{\emph{ECIS 2011 Proceedings}} (\bibinfo{date}{Oct.} \bibinfo{year}{2011}).
\newblock
\urldef\tempurl%
\url{https://aisel.aisnet.org/ecis2011/28}
\showURL{%
\tempurl}


\bibitem[Selbst et~al\mbox{.}(2019)]%
        {selbst_fairness_2019}
\bibfield{author}{\bibinfo{person}{Andrew~D. Selbst}, \bibinfo{person}{Danah Boyd}, \bibinfo{person}{Sorelle~A. Friedler}, \bibinfo{person}{Suresh Venkatasubramanian}, {and} \bibinfo{person}{Janet Vertesi}.} \bibinfo{year}{2019}\natexlab{}.
\newblock \showarticletitle{Fairness and {Abstraction} in {Sociotechnical} {Systems}}. In \bibinfo{booktitle}{\emph{Proceedings of the {Conference} on {Fairness}, {Accountability}, and {Transparency}}}. \bibinfo{publisher}{ACM}, \bibinfo{address}{Atlanta GA USA}, \bibinfo{pages}{59--68}.
\newblock
\showISBNx{978-1-4503-6125-5}
\urldef\tempurl%
\url{https://doi.org/10.1145/3287560.3287598}
\showDOI{\tempurl}


\bibitem[Strathern(1997)]%
        {strathern_improving_1997}
\bibfield{author}{\bibinfo{person}{Marilyn Strathern}.} \bibinfo{year}{1997}\natexlab{}.
\newblock \showarticletitle{‘{Improving} ratings’: audit in the {British} {University} system}.
\newblock \bibinfo{journal}{\emph{European Review}} \bibinfo{volume}{5}, \bibinfo{number}{3} (\bibinfo{date}{July} \bibinfo{year}{1997}), \bibinfo{pages}{305--321}.
\newblock
\showISSN{10627987, 1234981X}
\urldef\tempurl%
\url{https://doi.org/10.1002/(SICI)1234-981X(199707)5:3<305::AID-EURO184>3.0.CO;2-4}
\showDOI{\tempurl}


\bibitem[{The White House}(2023)]%
        {the_white_house_executive_2023}
\bibfield{author}{\bibinfo{person}{{The White House}}.} \bibinfo{year}{2023}\natexlab{}.
\newblock \bibinfo{title}{[{Executive} {Order} 14110 of {October} 30, 2023] {Safe}, {Secure}, and {Trustworthy} {Development} and {Use} of {Artificial} {Intelligence}}.
\newblock
\newblock
\urldef\tempurl%
\url{https://www.federalregister.gov/documents/2023/11/01/2023-24283/safe-secure-and-trustworthy-development-and-use-of-artificial-intelligence}
\showURL{%
\tempurl}


\bibitem[{US Department of Energy}(2022)]%
        {c2m2_us_department_of_energy_cybersecurity_2022}
\bibfield{author}{\bibinfo{person}{{US Department of Energy}}.} \bibinfo{year}{2022}\natexlab{}.
\newblock \bibinfo{booktitle}{\emph{Cybersecurity {Capability} {Maturity} {Model} ({C2M2})}}.
\newblock \bibinfo{type}{{T}echnical {R}eport} 2.1. \bibinfo{institution}{US Department of Energy}.
\newblock
\urldef\tempurl%
\url{https://www.energy.gov/ceser/cybersecurity-capability-maturity-model-c2m2}
\showURL{%
\tempurl}


\bibitem[Vakkuri et~al\mbox{.}(2021)]%
        {vakkuri_time_2021}
\bibfield{author}{\bibinfo{person}{Ville Vakkuri}, \bibinfo{person}{Marianna Jantunen}, \bibinfo{person}{Erika Halme}, \bibinfo{person}{Kai-Kristian Kemell}, \bibinfo{person}{Anh Nguyen-Duc}, \bibinfo{person}{Tommi Mikkonen}, {and} \bibinfo{person}{Pekka Abrahamsson}.} \bibinfo{year}{2021}\natexlab{}.
\newblock \bibinfo{title}{Time for {AI} ({Ethics}) {Maturity} {Model} {Is} {Now}}.
\newblock
\newblock
\urldef\tempurl%
\url{http://arxiv.org/abs/2101.12701}
\showURL{%
\tempurl}
\newblock
\shownote{arXiv:2101.12701 [cs]}.


\bibitem[Vorvoreanu et~al\mbox{.}(2023)]%
        {microsoft_vorvoreanu_responsible_2023}
\bibfield{author}{\bibinfo{person}{Mihaela Vorvoreanu}, \bibinfo{person}{Amy Heger}, \bibinfo{person}{Samir Passi}, \bibinfo{person}{Shipi Dhanorkar}, \bibinfo{person}{Zoe Kahn}, {and} \bibinfo{person}{Ruotong Wang}.} \bibinfo{year}{2023}\natexlab{}.
\newblock \bibinfo{booktitle}{\emph{Responsible {AI} {Maturity} {Model}}}.
\newblock \bibinfo{type}{{T}echnical {R}eport} MSR-TR-2023-26. \bibinfo{institution}{Microsoft}.
\newblock
\urldef\tempurl%
\url{https://www.microsoft.com/en-us/research/publication/responsible-ai-maturity-model/}
\showURL{%
\tempurl}


\bibitem[Weidinger et~al\mbox{.}(2023)]%
        {weidinger_sociotechnical_2023}
\bibfield{author}{\bibinfo{person}{Laura Weidinger}, \bibinfo{person}{Maribeth Rauh}, \bibinfo{person}{Nahema Marchal}, \bibinfo{person}{Arianna Manzini}, \bibinfo{person}{Lisa~Anne Hendricks}, \bibinfo{person}{Juan Mateos-Garcia}, \bibinfo{person}{Stevie Bergman}, \bibinfo{person}{Jackie Kay}, \bibinfo{person}{Conor Griffin}, \bibinfo{person}{Ben Bariach}, \bibinfo{person}{Iason Gabriel}, \bibinfo{person}{Verena Rieser}, {and} \bibinfo{person}{William Isaac}.} \bibinfo{year}{2023}\natexlab{}.
\newblock \bibinfo{title}{Sociotechnical {Safety} {Evaluation} of {Generative} {AI} {Systems}}.
\newblock
\newblock
\urldef\tempurl%
\url{http://arxiv.org/abs/2310.11986}
\showURL{%
\tempurl}
\newblock
\shownote{arXiv:2310.11986 [cs]}.


\bibitem[Wendler(2012)]%
        {wendler_maturity_2012}
\bibfield{author}{\bibinfo{person}{Roy Wendler}.} \bibinfo{year}{2012}\natexlab{}.
\newblock \showarticletitle{The maturity of maturity model research: {A} systematic mapping study}.
\newblock \bibinfo{journal}{\emph{Information and Software Technology}} \bibinfo{volume}{54}, \bibinfo{number}{12} (\bibinfo{date}{Dec.} \bibinfo{year}{2012}), \bibinfo{pages}{1317--1339}.
\newblock
\showISSN{09505849}
\urldef\tempurl%
\url{https://doi.org/10.1016/j.infsof.2012.07.007}
\showDOI{\tempurl}


\bibitem[Zhu et~al\mbox{.}(2021)]%
        {zhu_ai_2021}
\bibfield{author}{\bibinfo{person}{Liming Zhu}, \bibinfo{person}{Xiwei Xu}, \bibinfo{person}{Qinghua Lu}, \bibinfo{person}{Guido Governatori}, {and} \bibinfo{person}{Jon Whittle}.} \bibinfo{year}{2021}\natexlab{}.
\newblock \bibinfo{title}{{AI} and {Ethics} -- {Operationalising} {Responsible} {AI}}.
\newblock
\newblock
\urldef\tempurl%
\url{https://doi.org/10.48550/arXiv.2105.08867}
\showDOI{\tempurl}
\newblock
\shownote{arXiv:2105.08867 [cs]}.


\end{thebibliography}

\appendix
\section{Appendices}
\subsection{The Questionnaire}
\label{Questionniare-appendix}
In this appendix, we present the maturity model's questionnaire. As discussed in the body of the paper, the questionnaire is divided by stages of the development life-cycle. Each stage contains a number of topic statements, which contain individual statements. Evaluators can choose to evaluate topic statements only or all statements and are asked to provide both a score and evidence to support that score. Figure \ref{fig:questionnaire-structure} provides a visual illustration of the questionnaire's structure, and below is a list of all the statements. 

\hfill \break
\textbf{For AI at or after the planning and design stage:}
\begin{enumerate}
    \item Topic 1 - \emph{Mapping impacts:} We document what the AI will do and its potential impacts.
    \subitem{Substatements:}
    \begin{enumerate}
    \item{We document the \emph{goals, scope, and methods} of this AI system. (MAP 1.3, 2.1, 2.4, 3.3)}
    \item{We document the \emph{benefits and potential positive impacts} of this AI system, including the likelihood and magnitude. (MAP 1.1, 3.1, 5.1; GOV 4.2)}
    \item{We document the \emph{business value} of this AI system. (MAP 1.4, 3.1)}
    \item{We document the \emph{possible negative impacts} of this AI system, including the likelihood and magnitude. (GOV 1.1, 4.2, 5.1)}
    \item{We document the \emph{potential costs} of malfunctions of this AI system, including non-monetary costs such as decreased trustworthiness. (MAP 3.2)}
    \item{We implement processes to integrate input about \emph{unexpected impacts}. (MAP 5.2)}
    \item{We document the \emph{methods and tools} we use for mapping impacts. (MAP 2.3, 4.1)}
    \end{enumerate}
    \item Topic 2 - \emph{Documenting requirements:} We document basic requirements the system must meet
    \subitem{Substatements:}
    \begin{enumerate}
    \item{We document the \emph{human oversight} processes the system needs. (MAP 3.5)}
    \item{We document the \emph{technical standards and certifications} the system will need to satisfy. (MAP 3.4)}
    \item{We document AI \emph{legal requirements} that apply to this AI system. (GOV 1.1)}
    \end{enumerate}
    \item{Topic 3 - \emph{Culture:} We cultivate AI ethics mindsets}
    \subitem{Substatements}
    \begin{enumerate}
    \item{We write \emph{policies and guidelines} about AI ethics. (GOV 1.2, 1.4)}
    \item{We document \emph{roles, responsibilities, and lines of communication} related to AI risk management. (GOV 2.1)}
    \item{We provide \emph{training} about AI ethics to relevant personnel. (GOV 2.2)}
    \item{We implement practices to foster \emph{critical thinking} about AI risks. (GOV 4.1)}
    \end{enumerate}
\end{enumerate}

\hfill \break
\textbf{For AI at or after the model building and data collection stage:}
    \begin{enumerate} [(4)]
    \item{Topic 4 - \emph{Measuring risk:} We measure our potential negative impacts.}
    \begin{enumerate}
    \item{We document and periodically re-evaluate our \emph{strategy for measuring the impacts} of this AI system. It includes choosing which impacts we measure. It also includes how we will approach monitoring unexpected impacts and impacts that can't be captured with existing metrics. (MEA 1.1)}
    \item{We document the \emph{methods and tools} we use to measure the impacts of this AI system. It includes which metrics and datasets we use. MEA (2.1, 3.1, 3.2)}
    \item{We document the \emph{effectiveness of our measurement processes}. (MEA 2.13)}
    \item{We regularly evaluate and document the \emph{performance} of this AI system in conditions similar to deployment. (MEA 2.3)}
    \item{We regularly evaluate and document \emph{bias and fairness} issues related to this AI system. (MEA 2.11)}
    \item{We regularly evaluate and document \emph{privacy} issues related to this AI system. (MEA 2.10)}
    \item{We regularly evaluate and document \emph{environmental} impacts related to this AI system. (MEA 2.12)}
    \item{We regularly evaluate and document \emph{transparency and accountability} issues related to this AI system. (MEA 2.8)}
    \item{We regularly evaluate and document \emph{security and resilience} issues related to this AI system. (MEA 2.7)}
    \item{We regularly evaluate and document \emph{explainability} issues related to this AI system. (MEA 2.9)}
    \item{We regularly evaluate and document \emph{third-party issues, such as IP infringement}, related to this AI system. (MEA GOV 6.1)}
    \item{We regularly evaluate and document \emph{other impacts} related to this AI system. (We added on top of the RMF)}
    \item{If evaluations use \emph{human subjects}, they are representative and meet appropriate requirements. (MEA 2.2)}
    \end{enumerate} 
\end{enumerate}

\begin{enumerate} [(5)]
\item{Topic 5 - \emph{Transparency:} We document information about the system's limitations and risk control}
\subitem{Substatements:}
\begin{enumerate}
\item{We document information about the system's \emph{limitations and options for human oversight} related to this AI system. The documentation is good enough to assist those who need to make decisions based on the system's outputs. (MAP 2.2)}
\item{We document the system \emph{risk controls}, including in third-party components. (MAP 4.2)}
\item{We \emph{explain the model} to ensure responsible use. (MEA 2.9)}
\item{We inventory information about this AI system in a \emph{repository} of our AI systems. (GOV 1.6)}
\end{enumerate}
\end{enumerate}

\begin{enumerate} [(6)]
\item{Topic 6 - \emph{Management plan:} We plan how we will respond to risks }
\subitem{Substatements:}
\begin{enumerate}
    \item {We \emph{plan} and document how we will respond to the risks caused by this AI system. The response options can include mitigating, transferring, avoiding, or accepting risks. (MAN 1.3)}
    \item{We \emph{prioritize the responses} to the risks of this AI system based on impact, likelihood, available resources or methods, and the organization's risk tolerance. (MAN 1.2)}
    \item{We document the \emph{residual risks} of this AI system (the risks that we do not mitigate). The documentation includes risks to buyers and users of the system. (MAN 1.4)}
    \item{We have a plan for addressing \emph{unexpected risks} related to this AI system as they come up. (MAN 2.3)}
\end{enumerate}
\end{enumerate}

\begin{enumerate} [(7)]
\item{Topic 7 - \emph{Risk mitigation:} We act to minimize the risks we identify}
\subitem{Substatements:}
\begin{enumerate}
\item{We proactively evaluate whether this system \emph{meets its stated objectives} and whether its development or deployment should proceed. (MAN 1.1)}
\item{We ensure this AI's \emph{bias and fairness} performance stays meets our standards. (MAN 4.2)}
\item{We ensure this AI's \emph{privacy} performance meets our standards. (MAN 4.2)}
\item{We ensure this AI's \emph{environmental} performance meets our standards. (MAN 4.2)}
\item{We ensure this AI's \emph{transparency and accountability} meets our standards. (MAN 4.2)}
\item{We ensure this AI's \emph{security and resilience} meets our standards. (MAN 4.2)}
\item{We ensure this AI's \emph{explainability} performance meets our standards. (MAN 4.2)}
\item{We ensure this AI's \emph{third-party impacts, such as IP infringement}, meet our standards. (GOV 6.1)}
\item{We implement processes for \emph{human oversight} related to this AI system. (MAN 3.5)}
\item{We implement processes for \emph{appeal} related to this AI system. (MAN 4.1)}
\item{We maintain \emph{end-of-life} mechanisms to supersede, disengage, or deactivate this AI system if its performance or outcomes are inconsistent with the intended use. (MAN 2.4, GOV 1.6)}
\item{We address all \emph{other risks} prioritized in our plans related to this system by conducting measurable activities. (We added on top of the RMF)}
\item{We address \emph{unexpected risks} related to this system by conducting measurable activities. (MAN 2.3)}
\item{We track and respond to \emph{errors and incidents} related to this system by conducting measurable activities. (MAN 4.3)}
\end{enumerate}
\end{enumerate}

\hfill \break
\textbf{For AI at or after the deployment stage:}
\begin{enumerate} [(8)]
\item{Topic 8 - \emph{Pre-deployment checks:} We only release versions that meet our AI ethics standards}
\subitem{Substatments:}
\begin{enumerate}
    \item {We demonstrate that this system is \emph{valid, reliable, and meets our standards}. We document the conditions under which it falls short. (MEA 2.5; MAN 1.1)}
\end{enumerate}
\end{enumerate}

\begin{enumerate} [(9)]
\item{Topic 9 - \emph{Monitoring:} We monitor and resolve issues as they arise}
\subitem{Substatments:}
\begin{enumerate}
    \item {We \emph{plan} how to monitor risks related to this system post-deployment. (MAN 4.1)}
    \item{We monitor this system's \emph{functionality} and behavior post-deployment. (MEA 2.4)}
    \item{We apply mechanisms to \emph{sustain the value} of this AI system post-deployment. (MAN 2.2)}
    \item{We capture and evaluate \emph{input from users} about this system post-deployment. (MAN 4.1)}
    \item{We monitor \emph{appeal and override} processes related to this system post-deployment. (MAN 4.1)}
    \item{We monitor \emph{incidents} related to this system and responses to them post-deployment. (MAN 4.1)}
    \item{We monitor incidents related to \emph{high-risk third-party} components and respond to them. (GOV 6.2)}
    \item{We implement \emph{all other} components of our post-deployment monitoring plan for this system. (MAN 4.1)}
    \item{We monitor issues that would trigger our \emph{end-of-life} mechanisms for this system, and we take the system offline if issues come up. (MAN 2.4)}
\end{enumerate}
\end{enumerate}

\subsection {Scoring Examples}
\label{Scoring-examples-appendix}

In this appendix, we illustrate working with our maturity model by using it to evaluate governance based on the public disclosures of two companies: OpenAI and Duolingo. OpenAI develops generative AI chatbots. They detailed their ``Approach to AI Safety'' in \cite{openai_our_2023}, which was published in April 2023, at the release of GPT 4. As the name suggests, the document presents the company’s AI ethics approach. Dulingo’s most prominent product is an app for learning languages. The Duolingo English Test is an English proficiency test intended to be used as a university entry requirement for non-native English speakers, similar to the TOEFL exam. Dulingo uses generative AI in the question-writing process and in the evaluation process. The ``English Test Responsible AI Standards'' \cite{burstein_jill_duolingo_2023} is intended to guide their research and documentation.

For the sake of the example, we will use these documents as the sole source of evidence for scoring one topic: ''Measuring risk - We measure our potential negative impacts”. The ''Measuring Risk” topic requires the company to evaluate and document the impact in the following areas regularly: performance, bias and fairness, privacy, environmental impact, transparency and accountability, security and resilience, explainability, and third-party issues such as IP infringement. We add to these NIST requirements that the company regularly evaluate and document other relevant impacts, even if they were not listed by NIST. In addition, ``Measuring Risk'' requirements also include evaluation of the measurement process itself, which means documenting and regularly re-evaluating the strategy for impact measurement, the methods and tools used for measurement, and the effectiveness of the measurement process. Last, ``Measuring Risk'' also requires that if the evaluations use human subjects, they are representative and meet appropriate requirements.

We discuss evidence relevant to the scoring of ``Measuring Risk'' topic and assess the degree to which the three metrics are satisfied. This discussion reflects the results of our deliberations and it is intended to illustrate our way of thinking.

\subsubsection{Coverage of NIST’s recommended activities}

OpenAI’s ``Approach of AI Safety'' \cite{openai_our_2023} is a 1200-word document that covers the following topics: Building increasingly safe AI systems; learning from real-world use to improve safeguards; Protecting children; Respecting privacy; and Improving factual accuracy. Each of these topics is addressed in a few paragraphs that describe the company’s approach efforts related to that topic. 

The document presents a partial coverage of the risk area NIST specifies. Only performance and privacy are mentioned. Several of the other risk areas that NIST mentions are omitted despite their centrality to OpenAI’s product. For example, the lack of attention to IP infringement and fairness issues is especially striking. However, OpenAI’s document adds a topic not specified by NIST: protecting children. Therefore, the document satisfies the “coverage” metric to a low degree. 

Duolingo’s The ``English Test Responsible AI Standards'' \cite{burstein_jill_duolingo_2023} is an 18-page document intended to guide their research and documentation. It covers the following topics: Validity and reliability; Fairness; Privacy and security; and Accountability and transparency. Each of these topics includes a statement of the company’s goals related to that topic and details on the processes they implement to achieve these goals. 

Duolingo’s document covers most of the risk areas NIST specifies, as it addresses performance, fairness, privacy, transparency and accountability, explainability (as part of other sections), and security and resilience are covered. The topics that are missing are environmental impacts and third-party issues (e.g., copyright issues stemming from the use of pre-trained models). Therefore, it satisfies the “coverage” metric to a high degree. 

\subsubsection{Robustness}
As discussed above, robustness covers six ideals for the AI risk management activities: (i) Regularity - Performed in a routine manner; (ii) Systematicity - Follow policies that are well-defined and span company-wide; (iii) Trained Personnel - Performed by people who are properly trained and whose roles in the activities are clearly defined; (iv) Sufficient Resources - Supported by sufficient resources, including budget, time, compute power, and cutting-edge tools; (v) Adaptivity - Adapting to changes in the landscape and product, including regular reviews and effective contingency processes to respond to failure; and (iv) Cross-functionality - Involve all core business units and senior management. They are informed of the outcomes and contribute to decision-making, strategy, and resource allocation related to the activities. 

OpenAI’s document satisfies the robustness metric to a low degree because the description of the relevant processes lacks detail that would illustrate satisfaction of the robustness ideals. For example, the section about accuracy describes the efforts in one sentence only, this sentence:
\begin{quote}
By leveraging user feedback on ChatGPT outputs that were flagged as incorrect as a main source of data—we have improved the factual accuracy of GPT-4. GPT-4 is 40\% more likely to produce factual content than GPT-3.5.    
\end{quote}

This sentence describes an end result of decreasing inaccuracy by 40\%. However, the document doesn’t address topics such as what the level of inaccuracy is now, what it used to be, how exactly inaccuracy is quantified, or what practices the company implements to regularly monitor its accuracy (if at all). These details are crucial for determining whether accuracy-related processes satisfy the robustness ideals. 

At the top of the document, OpenAI states that
[A]fter our latest model, GPT-4, finished training, we spent more than 6 months working across the organization to make it safer and more aligned prior to releasing it publicly.
This sentence suggests some robustness. However, the document provides no further descriptions or explanations that would indicate that this sentence is more than a platitude. 

The lack of detail in OpenAI’s “AI Safety Approach” is especially striking in comparison to their ''Preparedness Framework (Beta)”, which came out in December 2023 \cite{openai_preparedness_nodate}. This document describes their approach to risk management with regards to what they call ``catastrophic risk'', which they define to be ``any risk which could result in hundreds of billions of dollars in economic damage or lead to the severe harm or death of many individuals —this includes, but is not limited to, existential risk'' (p. 2). The ``Preparedness'' document is very detailed. Over its 27 pages, it describes topics such as how they will use evaluations to track catastrophic risks, continuously identify unknown categories of catastrophic risks, what the ``Preparedness'' team will do, and the cross-functionality of the advisory board. This document suggests plans for high degree of robustness, but only regarding one risk area, catastrophic risk. It highlights the fact that the ``AI Safety'' document lacks similar evidence for robustness with regard to the risk areas recommended by the NIST AI RMF, such as fairness.

In comparison, Duolingo’s document presents much stronger evidence for robustness. For example, the section about validity and reliability states that the rationale of these standards is ''to ensure that the test is suitable for its intended purpose” and then goes on to specify goals their work processes aim to achieve as well as recommended activities. 

The level of detail in this description makes the document stronger evidence that the company’s efforts to ensure good AI performance are robust. They are well thought out, systematic, and lend themselves to implementation. However, the evidence would be stronger if the document included explanations about how these processes are integrated into the organization’s day-to-day work or other indications that they are indeed implemented. Therefore, we see the document as providing medium evidence for the robustness of the “Mitigating Risk” topic. 

\subsubsection{Input Diversity}

The input metric is about the extent to which the company integrates input from diverse stakeholders in its processes. OpenAI’s document mentions user feedback in one sentence in the ''Learning from real-world examples” section:
We cautiously and gradually release new AI systems—with substantial safeguards in place—to a steadily broadening group of people and make continuous improvements based on the lessons we learn. 

However, it is unclear to what extent the feedback is diverse and to what extent it is adopted. Therefore, the document satisfies the ''input diversity” metric to a low degree. 

Duolingo’s document mentions details that are relevant to input diversity. For example, the document requires that employees ''Evaluate and document demographic representation in data sets used to build AI” \cite{burstein_jill_duolingo_2023}. However, key sections lack an indication that the company collects and uses feedback from diverse internal and external stakeholders. For example, Goal 4.1 is to ``Assess how AI processes impact stakeholders'' \cite{burstein_jill_duolingo_2023}. The requirements include processes to document impacts and risks but not processes for direct engagement with stakeholders such as test takers and universities. Therefore, Duolingo’s document also satisfies the ``input diversity'' metric to a low degree. 

\subsubsection{Overall score for the \emph{Measuring Risk} topic}
Recall that the scoring question is whether ``There is evidence that the company performs the relevant activities in a way that satisfies all metrics to a high degree'', and evaluators are asked to indicate whether they strongly agree (4), somewhat agree (3), somewhat disagree (2), or strongly disagree (1). 

OpenAI’s document exhibits low satisfaction of all three metrics, coverage, robustenss, and input diversity, with regard to the ``Measuring Risk'' topic. Therefore, we determine that they deserve the score of ``1'' for this topic (based on this document). 

Duloingo’s document exhibits variability: high coverage, medium robustness, and low input diversity. Therefore, we determine that they deserve a score of ``3'' for the risk measurement topic (based on this document).

\end{document}